\documentclass[preprint,3p,times,numbers]{elsarticle}
\usepackage{amsmath,booktabs,graphicx,natbib,array,caption,multirow}
\usepackage[hidelinks]{hyperref}
\usepackage{adjustbox}
\usepackage{xurl}
\usepackage{xcolor}
\journal{Applied Energy}

\begin{document}

\begin{frontmatter}

\title{Price co-movement and price information diverge across Nordic electricity zones after the energy crisis}

\author[1]{My Thi Diem Phan\corref{cor1}}
\author[2]{Tuyen Trung Truong}
\author[1]{Hoai Phuong Ha}
\author[3]{Dat Thanh Nguyen\corref{cor1}}
\address[1]{Department of Computer Science, UiT The Arctic University of Norway, P.O. Box 6050 Langnes, 9037 Troms{\o}, Norway}
\address[2]{Department of Mathematics, University of Oslo, P.O. Box 1053 Blindern, 0316 Oslo, Norway}
\address[3]{Faculty of Medicine, University of Oslo, P.O. Box 1078 Blindern, 0316 Oslo, Norway}

\cortext[cor1]{Corresponding authors. Email addresses diemmyphan@outlook.com and thanhdng@uio.no.}

\begin{abstract}
After the 2021-2022 energy crisis, day-ahead prices in Nordic bidding zones no longer move together. We ask whether this divergence also changes the information carried by prices. The answer matters for integrating variable renewable energy (VRE), because it can help identify where wind and solar forecasts are worth producing, publishing, and using. We measured the co-movement of each Norwegian and Swedish zone with Germany-Luxembourg (DE-LU) in 2018-2025. We then used a single model, selected on 2024 data, to forecast 2025 prices at 11:00 on the previous day and tested which inputs improved accuracy in each zone. The zones moved alike until 2019, but co-movement diverged in Sweden in 2020 and in Norway in 2021, and the split has persisted. In 2024, four southern zones co-moved strongly with DE-LU (intraday correlations of 0.54-0.71), while four northern zones co-moved weakly (0.25-0.30). Yet price information did not follow this split. Previous-day neighbouring prices reduced mean absolute error by at most 2.2\% in any zone, equally in coupled and isolated zones. Removing the transmission system operators' wind and solar forecasts increased mean absolute error by at most 6\% in Norway, 8-16\% in Sweden, and 31\% in DE-LU, and the ordering held in 2024. Within Norway, the value of these forecasts showed no detectable dependence on reservoir filling. The divergence in co-movement was therefore not matched by a divergence in price information, which differed mainly between national generation systems. Thus, a zone's coupling to the continent does not indicate the value of VRE forecasts there.
\end{abstract}


\end{frontmatter}

\section{Introduction}
\label{sec:intro}
\begin{figure}[!tbp]
\centering
\includegraphics[width=\textwidth]{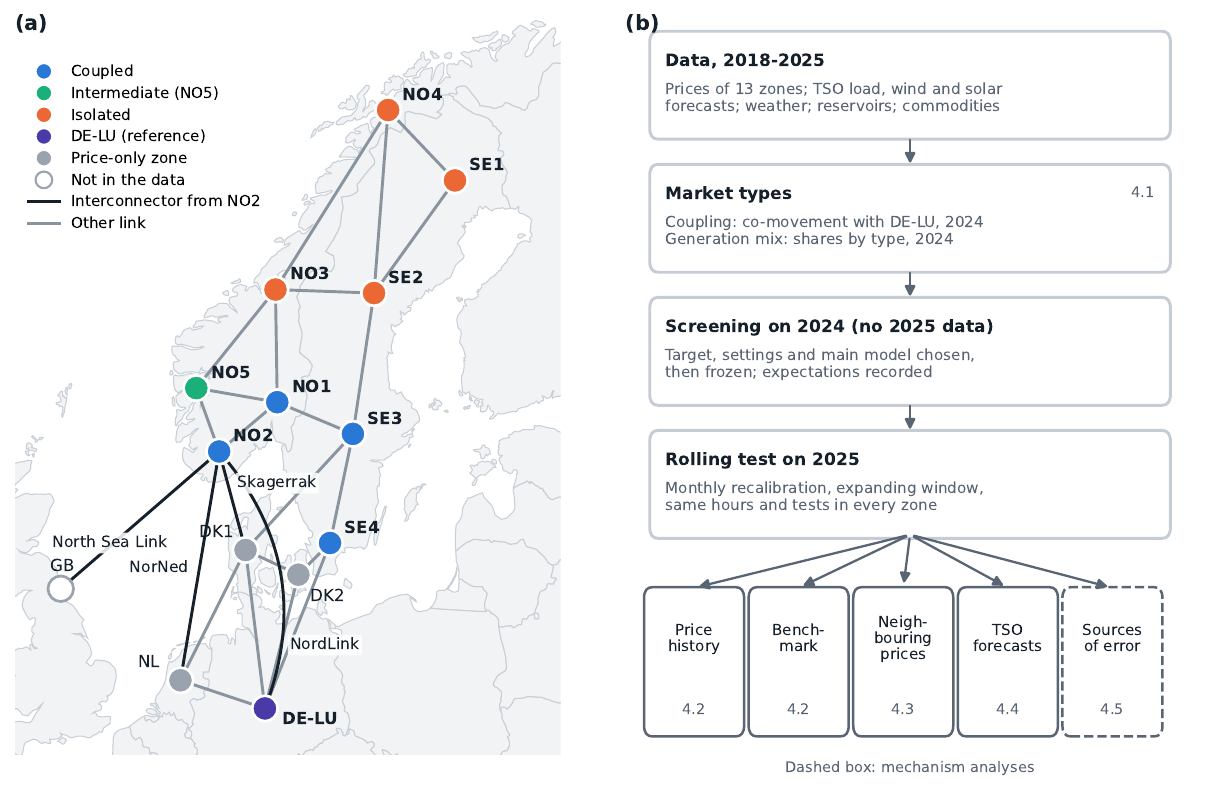}
\caption{Bidding zones and study design.
(a) The zones as a network.
Nodes are placed by hand at approximate zone locations and coloured by market type from the 2024 coupling measure.
Grey nodes are price-only zones, and Great Britain (open) is not in the data.
All links are drawn with the same width.
The four named interconnectors from NO2 are NordLink to DE-LU and North Sea Link to Great Britain (1,400~MW each) \citep{Statnett2021nordlink}, NorNed to NL (700~MW) \citep{NordPool2013ntc} and Skagerrak to DK1 (about 1,600~MW) \citep{StatnettEnerginet2019skagerrak}.
NordLink is drawn curved only for legibility.
The other lines mark neighbouring areas with a direct link \citep{nordictso2023}. Links to Finland, the Baltic states and Poland are not shown. Map lines delineate study areas and do not necessarily depict accepted national boundaries.
Base map from Natural Earth, 1:50m countries (public domain).
(b) Study design, with the section of each experiment.}
\label{fig:design}
\end{figure}

European electricity systems are changing in two ways at the same time. First, bidding zones are increasingly linked by interconnectors and cleared together through day-ahead market coupling, and price convergence across zones is used as an indicator of market integration~\citep{ACER2022MMR}. Second, the share of wind and solar power, or variable renewable energy (VRE), is growing. Wind and solar output lowers the price level and raises price volatility, and this effect grows with the installed capacity~\citep{hirth2013,ketterer2014}. As a result, prices depend more and more on weather-driven supply. The Nordic market combines both changes in an unusual way. Norway and northern Sweden rely mainly on hydropower, whereas southern Sweden relies mainly on nuclear and wind power, and the direct links to the continent and Great Britain originate from NO2 in Norway and from SE3 and SE4 in Sweden (Fig.~\ref{fig:design}a). Between 2019 and 2022, this setting changed quickly. Two new interconnectors linked southern Norway to Germany and Great Britain~\citep{Statnett2021nordlink,Statnett2021nsl}, Sweden closed two nuclear reactors in SE3~\citep{holmberg2023}, and record gas prices in 2022 caused large price differences between European regions~\citep{ACER2022MMR}.

These two changes raise two practical questions. The first concerns interconnection, which determines how far continental shocks, such as gas prices and German wind and solar output, reach Nordic producers and consumers. The second concerns the local generation mix, which determines how far local wind and solar variability reaches prices and therefore which information is useful for forecasting them. In DK1, day-ahead wind power forecasts lower spot prices and change their distribution~\citep{jonsson2010}. The second question also has a direct operational side. Transmission system operators (TSOs) produce wind and solar forecasts, but European rules require them to publish these forecasts only by 18:00 on the day before delivery, which is after the day-ahead auction has closed~\citep{eu543}. Whether earlier publication would help market participants therefore depends on how much information these forecasts add to prices in each zone. If the value of VRE forecasts followed coupling, the zones linked to the continent would need them most. If it follows the generation mix, the priorities are different.

Existing research has documented the rise in southern Nordic prices and several of its drivers and consequences. Holmberg and Tanger\aa s~\citep{holmberg2023} report that Swedish zonal prices were similar until 2020, after which grid bottlenecks insulated the northern zones from the energy crisis. Mauritzen~\citep{mauritzen2025great} and Brolinson et al.~\citep{brolinson2026} use the Norwegian price gap as a natural experiment to study its wider economic effects. Bj{\o}rndal et al.~\citep{bjorndal2026} find that NordLink raised NO2 prices in expensive hours, lowered them in cheap hours and strengthened their exposure to gas prices, while Bendiksen et al.~\citep{bendiksen2025} associate cross-border flows with the tails of the NO2 price distribution but find that these flows do not improve out-of-sample forecasts of that distribution. Further studies trace the transmission of gas prices to Nordic prices~\citep{zhu2024}, the response of North Sea flows to price differences~\citep{saether2024effect} and the market value of Norwegian hydropower~\citep{tselika2025}. Trebbien et al.~\citep{trebbien2024} analysed the correlations of European day-ahead prices from 2019 to 2023 and found that correlations within the Nordic region fell during the crisis, while the southern Norwegian zones joined a community of correlated continental zones. These studies compare the periods before and during the crisis, or they examine price levels, distributions or single drivers in one zone or pair of zones. To our knowledge, none has measured the co-movement of each Nordic zone with continental prices year by year beyond the crisis, or asked whether the information that prices carry follows the same split.

Answering these questions is harder than comparing price levels. All zones share seasonal and crisis-driven swings, so prices at very different levels may still move together, and similar levels need not mean similar movements. The relevant measure is therefore co-movement across zones, years and time scales. Measuring the information that prices carry is harder still. In the forecasting literature, information from connected markets~\citep{lago2018integration,chomon2026,mascarenhas2026} and fundamental inputs are mainly used to improve accuracy~\citep{lago2021}, not to compare what prices in different systems contain. Such a comparison requires a strict information set, so that every input is known at the time of the forecast, and one forecasting approach that works in every zone, so that differences reflect the markets rather than the models. It must also cope with the regime break of the crisis, which makes older data look unlike recent prices~\citep{marcjasz2018calibration,uniejewski2026}, and fix its modelling choices before the test data are examined.

We address three research questions. First, has the Nordic market split into groups of zones with different price movements, and when did the split occur? Second, does the information that prices carry follow this split, so that prices of neighbouring markets are more useful in the zones coupled to the continent? Third, does this information instead follow the generation mix, so that VRE forecasts are more useful where wind and solar make up more of local supply? For the first question, we measure the correlation of every Norwegian and Swedish zone with Germany-Luxembourg (DE-LU) for each year from 2018 to 2025, at three time scales. For the second and third questions, we forecast all 24 hourly prices of each day of 2025 at 11:00 on the preceding day, using only information available at that time. We select one model, per-hour LightGBM~\citep{ke2017lightgbm} with prices normalised by the previous day's mean and spread (Eq.~\ref{eq:norm}), on 2024 data and fix it before the 2025 test. We then add or remove inputs, that is, explanatory variables of the forecasting model, such as the previous-day prices of neighbouring markets, the TSO wind and solar forecasts and reservoir levels. Following the predictive notion of Granger~\citep{granger1969}, we read a larger out-of-sample gain from an input as evidence that the prices of that zone contain less of the information in that input. Finally, we relate the gain from VRE forecasts to the generation mix across zones and to reservoir filling within Norwegian zones, and we use double machine learning~\citep{chernozhukov2018} to examine on which days prices are easier or harder to forecast.

The results show that co-movement and price information divide the zones in different ways. Co-movement with DE-LU was similar across zones in 2018-2019, diverged in Sweden in 2020 and in Norway in 2021, and has persisted beyond the crisis. Four southern zones (NO1, NO2, SE3 and SE4) remain coupled to continental prices, four northern zones (NO3, NO4, SE1 and SE2) are isolated from them, and NO5 lies between the two groups. The information carried by prices, however, does not follow this split. Neighbouring prices add little in both groups, whereas VRE forecasts add little in Norway and much in Sweden and DE-LU, including in Swedish zones whose wind shares are close to those of Norwegian zones. The same ordering appears when the comparison is repeated on 2024 data (rank correlation across zones 0.93; Supplementary Section~S10). These results rest on one test year, point forecasts and ten zones, in which the country and the generation mix cannot be fully separated, and they assume that VRE forecasts are available at 11:00 on the day before delivery (Section~\ref{sec:d_limits}).

We make three contributions:
\begin{itemize}
\item We show that the Nordic split is a persistent change in co-movement, not only a gap in price levels. In 2024, the correlation of the intraday price profile with DE-LU was 0.54-0.71 in the coupled zones and 0.25-0.30 in the isolated zones, with NO5 at 0.42 (Section~\ref{sec:r_types}).
\item We show that the information prices carry does not follow coupling. Previous-day neighbouring prices lower the relative mean absolute error (rMAE) by at most 0.0095 in any zone, and gains larger than about 0.02 are excluded in every zone. The gain in coupled zones exceeds that in isolated zones by only 0.0004 (95\% interval $-0.0055$ to $0.0061$; Section~\ref{sec:r_neighbours}).
\item We show that the value of VRE forecasts differs instead between national generation systems. Removing them raises rMAE by 0.012-0.032 in NO1-NO4, by 0.036-0.073 in Sweden and by 0.130 in DE-LU, or by 0.3-0.5, 0.8-2.8 and 4.3~EUR/MWh. This value shows no detectable dependence on reservoir filling within Norwegian zones, and in the isolated zones prices are easier to forecast on days with high wind and solar output (Sections~\ref{sec:r_tso} and~\ref{sec:r_errors}).
\end{itemize}
A common forecasting approach makes these comparisons possible. With the normalised target, per-hour LightGBM uses the full price history across the regime break, reduces the MAE of the D-1 naive forecast by 24-37\% in the Nordic zones, and is not significantly beaten by any other model in any zone (Section~\ref{sec:r_history}).

The rest of the paper is organised as follows. Section~\ref{sec:data} introduces the market, data and information set, and Section~\ref{sec:methods} describes the methods. Section~\ref{sec:results} presents the results in the order of the three research questions. Section~\ref{sec:discussion} discusses what they imply for price formation, VRE forecasting and market practice, and Section~\ref{sec:conclusion} concludes.

\section{Market, data and information set}
\label{sec:data}

\subsection{Market and zones}
\label{sec:d_market}
We model the five Norwegian price zones (NO1-NO5), the four Swedish zones (SE1-SE4) and DE-LU, which serves as the continental reference and whose price before October 2018 corresponds to the joint German-Austrian-Luxembourg zone (Fig.~\ref{fig:design}a). DK1, DK2 and NL are included only as price series. Their prices enter the co-movement analysis, and those of DK1 and NL, together with DE-LU, also serve as neighbouring prices (Section~\ref{sec:m_neighbours}). Norway is connected to continental Europe and Great Britain through four interconnectors, all from NO2. These are Skagerrak to DK1, which consists of four cables \citep{StatnettEnerginet2019skagerrak}, NorNed to NL \citep{NordPool2013ntc}, NordLink to Germany, which began trial operation on 9 December 2020 \citep{Statnett2021nordlink}, and North Sea Link to Great Britain, which began trial operation on 1 October 2021 \citep{Statnett2021nsl}. Sweden, in contrast, is connected to the continent from two zones, to DK1 from SE3 through Konti-Skan \citep{svk_kontiskan} and to DK2 and Germany from SE4 through the {\"O}resund link and the Baltic Cable \citep{nordictso2023}. Within and between the two countries, the zones are linked by internal transmission lines with limited capacity.

All these zones are cleared in the Nord Pool day-ahead auction, which closes at 12:00 CET on the day before delivery (D-1) and sets the prices for all hours of the delivery day (D) at once \citep{nordpool_da,entsoe_sdac}. We therefore forecast all 24 hourly prices of D at 11:00 (Europe/Oslo) on D-1, one hour before the auction closes, when the prices of D-1 are already known from the auction on D-2 \citep{lago2021}. Two recent changes to the auction affect our data. Nordic flow-based market coupling began on 29 October 2024, with 30 October as the first delivery day \citep{statnett2024fb}, and from 1 October 2025 day-ahead prices have been set for 15-minute intervals instead of hours \citep{sdac2025,entsoe_sdac}, so we average the four quarter-hourly prices to obtain hourly prices.

\subsection{Data}
\label{sec:d_data}
The data cover delivery days from 8 January 2018 to 31 December 2025, with one earlier month kept as history for the first forecasts.
Day-ahead prices, the load forecasts and the wind and solar forecasts of the TSOs (Table~\ref{tab:inputs}), the weekly stored energy in hydro reservoirs (Norway and Sweden) and the actual generation by production type (2024, for Table~\ref{tab:types} only) come from the Transparency Platform of the European Network of Transmission System Operators for Electricity (ENTSO-E) \citep{entsoe_tp,eu543}.
For Norway, the weekly fill share of the reservoirs and its 20-year median for the same week come from the Norwegian Water Resources and Energy Directorate (NVE) \citep{nve_magasin}. They define the reservoir anomaly of Section~\ref{sec:m_errors} and are never model inputs.
Hourly weather comes from the Open-Meteo historical archive \citep{openmeteo}, which combines ECMWF IFS data with the ERA5 and ERA5-Land reanalyses \citep{hersbach2020era5}, at one point in each Norwegian zone, one or two points in each Swedish zone and six points in DE-LU, averaged within each zone.
Daily closes of the front-month TTF gas, API2 coal, Brent oil and Henry Hub gas futures and of the EUR/NOK exchange rate come from Yahoo Finance \citep{yahoo}.
The carbon price is the daily price of the front-December EU ETS allowance (EUA) future in EUR/t, from Ember based on Montel data \citep{ember_prices}.
Coal is converted from USD/t to EUR/t with the daily ECB reference rate \citep{ecb_fx}.
The TTF and coal series switch to the next contract every month, which creates a jump on the first trading day after each expiry. Their returns, volatility and momentum are therefore computed from a ratio roll-adjusted series, in which the return on that day is set to zero, while the price levels are used as quoted.

All ten modelled zones have all hours of the period and no missing prices after processing.
We checked the price processing independently by re-deriving the hourly prices of all 13 zones from the raw ENTSO-E files.
The check covered the time resolution, the hourly averaging, gaps and the daylight-saving mapping, and every check passed (Supplementary Section~S1).
Forecast series that are missing in more than 5\% of the hours of a zone are dropped for that zone, which removes offshore wind in Norway.
The wind and solar forecast of NO5 is zero in every year, in line with its generation mix (Table~\ref{tab:types}).
The price of 31 December 2025 at 00:00 is the mean of three quarter-hours instead of four in all zones, because the full-year ENTSO-E request omits the first value of that day.
We retain and document this discrepancy rather than impute the missing value. It affects one test hour per zone.

\subsection{Information set}
\label{sec:infoset}
Every input is dated by the time its measurement is complete, and a forecast for D uses only values whose event or measurement took place before 11:00 on D-1 (Table~\ref{tab:inputs}).
A weekly reservoir value, for example, is dated at the end of the week it describes.
The weather data are reanalysis observations rather than forecasts, so only observations up to 10:00 on D-1 are used.
Commodity closes of D-1 are known only in the afternoon, so the latest usable close is that of D-2.

The TSO forecasts for D are the one exception that depends on publication time.
The load forecast must be published no later than two hours before gate closure, but the wind and solar forecasts must be published only by 18:00 Brussels time on D-1 \citep{eu543}.
The raw data contain no publication times, so we cannot tell whether a given wind and solar forecast was available at 11:00.
We assume zero publication delay for all inputs.
Section~\ref{sec:r_tso} reports how much accuracy depends on this assumption.

Days with 23 or 25 hours, which arise from the daylight-saving changes, are mapped to 24 hourly slots \citep{lago2021}.
On 23-hour days, the missing 02:00 slot is filled with the mean of the prices at 01:00 and 03:00 and is excluded from all error measures.
On 25-hour days, the two prices at 02:00 are averaged.
Short gaps in the input series are filled with the last known value.
A data-quality check in preprocessing allows at most five weekdays of forward filling or of identical values in a daily series and at most 120 hours in an hourly input. Runs of identical values in slow weather variables (snow depth, soil moisture and soil temperature) and in cloud cover at 0 or 100\% are reported but allowed.

\begin{table}[!tbp]
\centering
\small
\caption{Model inputs, the values used for delivery day D at the forecast time of 11:00 on D-1, and the models that use them.
L denotes LEAR, H per-hour LightGBM and P pooled LightGBM.
$L_D$ and $S_D$ are the mean and standard deviation of the 24 prices of D-1 (Section~\ref{sec:m_norm}).}
\label{tab:inputs}
\begin{tabular}{>{\raggedright\arraybackslash}p{2.4cm}>{\raggedright\arraybackslash}p{5.0cm}>{\raggedright\arraybackslash}p{5.4cm}>{\raggedright\arraybackslash}p{1.4cm}}
\toprule
Group & Variables & Values used for D & Models \\
\midrule
Prices & Day-ahead price of the zone & All 24 hours of D-1, D-2, D-3 and D-7; $L_D$ and $S_D$ for H and P & L, H, P \\
Load & TSO load forecast & Each hour of D, and of D-1 and D-7 for L and H & L, H, P \\
Wind and solar & TSO forecasts of onshore wind and solar generation, and their total & Each hour of D, and of D-1 and D-7 for L and H & L, H, P \\
Weather & 20 variables, including temperature, wind, radiation, precipitation and snow & Observation at 10:00 on D-1 and mean temperature from 00:00 to 10:00 on D-1 & P \\
Reservoir & Stored energy and its change over one and four weeks (Norway and Sweden) & Latest weekly value complete at 11:00 on D-1 & P \\
Commodities & TTF gas, EUA carbon (EUR/t), API2 coal (EUR/t), Brent oil, Henry Hub gas, EUR/NOK and derived features & Close of D-2 & P \\
Calendar & Weekday (L, H); hour, weekday and month (P) & Of D & L, H, P \\
Neighbouring prices & Day-ahead prices of DE-LU, DK1 and NL, normalised by their own level and spread, and three level gaps & All 24 hours of D-1 (Section~\ref{sec:r_neighbours} only) & H \\
\bottomrule
\end{tabular}
\end{table}

\section{Methods}
\label{sec:methods}

We first characterised the different market types and then used a fixed set of forecasting experiments to examine what information prices contain in each type (Fig.~\ref{fig:design}). All models were selected using data from 2024, before any forecasts for 2025 were made. Every experiment in 2025 followed the same monthly rolling scheme.

\subsection{Measuring market types}
\label{sec:m_types}
We characterised market coupling by the co-movement of each zone's prices with DE-LU prices, rather than by differences in price levels. For each zone and calendar year from 2018 to 2025, we computed the correlation of hourly day-ahead prices with DE-LU. Because correlations of price levels can be dominated by common seasonal and crisis-driven swings, we repeated the calculation at two shorter time scales. The intraday profile is the hourly price minus its daily mean, while the day-to-day movement is the daily mean price minus its calendar-month mean. For the supplementary figures, we also clustered the zones separately for each year using average linkage on $1-r$, with optimal leaf ordering \citep{barjoseph2001}.

After the crisis, the zones form two clearly separated groups, although NO5 lies between them from 2024 onward and coupling varies within each group (Section~\ref{sec:r_types}). We therefore used a continuous coupling measure in all tests. It is the 2024 correlation with DE-LU of hourly prices after subtracting each day's mean, calculated for each of the nine Norwegian and Swedish zones before any 2025 forecasts were made (Table~\ref{tab:types}). This intraday measure removes seasonal and crisis-driven level swings and captures the shape of daily prices that day-ahead forecasts must predict. Correlations of hourly price levels and day-to-day movements are reported for comparison. Two alternative measures, the share of hours in 2024 when a zone's price was within 1~EUR/MWh of the DE-LU price and the correlation with DE-LU of hour-to-hour price changes, are reported in Table~S4 but were not used in any test.

The continuous measure ranges from 0.248 in NO4 to 0.710 in SE4, with a median of 0.424 (NO5) and a mean of 0.441. For descriptive purposes, we call the zones above the median coupled (NO1, NO2, SE3, SE4) and those below it isolated (NO3, NO4, SE1, SE2), with NO5 classified as intermediate. The continuous measure retains variation within these groups and does not force NO5 into either. These labels reflect measured price co-movement rather than physical interconnection. For example, NO1 is linked to SE3 rather than directly to the continent, yet its 2024 coupling measure (0.551) exceeds the median.

The second dimension of the market types is the generation mix. Norway is dominated by hydropower, while Sweden has a mix of hydropower, nuclear and wind power. DE-LU serves as the continental reference. We calculated each zone's 2024 generation shares by production type using ENTSO-E actual generation data \citep{entsoe_tp} (Table~\ref{tab:types}).

\subsection{Evaluation design}
\label{sec:m_design}
The evaluation had two stages, a screening on 2024 and a monthly rolling test on 2025.
In the screening, no 2025 data were loaded.
LightGBM and the neural networks were fitted on delivery days from 8 January 2018 to 31 December 2022, stopped early on 2023, and refitted on 2018-2023 with the selected number of trees or epochs.
LEAR chose its penalty by the Akaike information criterion on 2018-2023.
All models were evaluated on 2024.
Target versions and LightGBM settings were chosen on the mean 2024 MAE over NO1-NO5, and the main model was the one with the lowest mean 2024 MAE over NO1-NO5.
These choices were then frozen for all experiments on 2025 (Table~S1).

In the test on 2025, every model was recalibrated at the start of each month.
The training data for month $m$ end on the last day before $m$, so no data from the test month enter training.
For LightGBM, the number of trees was chosen by early stopping on the last part of the training window and the model was then refitted on the whole window.
LEAR was recalibrated in the same way, with its penalty chosen by the Akaike information criterion on the whole training window, and the naive forecasts need no fitting.
The early-stopping block is one quarter of the window, at most 12 months, so it is 12 months for the expanding window and 3, 6 and 9 months for windows of one, two and three years.
The expanding window starts on 8 January 2018 and is the default in all experiments.

\subsection{Target normalisation}
\label{sec:m_norm}
Prices changed by an order of magnitude between 2018 and 2025, so we also expressed the target relative to the previous day.
Let $L_D$ and $S_D$ be the mean and the standard deviation of the 24 prices of D-1, with $S_D$ floored at 1~EUR/MWh.
The level-and-scale target is
\begin{equation}
y_{D,h} = \frac{p_{D,h} - L_D}{S_D}, \qquad h = 0, \dots, 23,
\label{eq:norm}
\end{equation}
where all lagged price inputs are transformed in the same way and $L_D$ and $S_D$ are added as inputs.
A prediction is converted back as $\hat p_{D,h} = L_D + S_D\,\hat y_{D,h}$.
The level-only target sets $S_D = 1$, and the raw target uses the price itself.
The idea follows reversible instance normalisation \citep{kim2022revin} and variance-stabilising transformations \citep{uniejewski2018,uniejewski2026}, and we do not propose it as a new method.

\subsection{Models}
\label{sec:m_models}
We compared naive forecasts, a linear benchmark, two gradient-boosting layouts and their ensemble.
The naive forecasts repeat the price of the same hour on D-1 or on D-7 \citep{lago2021}.
LEAR fits one LASSO regression \citep{tibshirani1996} per hour of D on 247 inputs \citep{lago2021}.
They are the 24 prices of D-1, D-2, D-3 and D-7, the 24 hourly load forecasts and the 24 hourly wind and solar forecasts for D, D-1 and D-7, and seven weekday dummies.
Inputs and target are scaled by their median and median absolute deviation and transformed with the inverse hyperbolic sine \citep{uniejewski2018,ziel2018}.
The penalty is chosen per hour by the Akaike information criterion on the training data, so no validation set is needed \citep{zou2007}.

LightGBM \citep{ke2017lightgbm} was used in two layouts.
The per-hour layout fits 24 models, one per hour of D, on the 247 LEAR inputs plus $L_D$ and $S_D$.
The pooled layout fits one model on one row per delivery hour with 88 inputs, which add weather, reservoir and commodity variables to the price, forecast and calendar inputs.
The commodity variables are, for each series, the close of D-2, its logarithm, its 5- and 20-day means, its daily return and the 20-day standard deviation of the return, together with the ratios of the TTF price to the carbon, Brent and coal prices and the momentum of gas and carbon, defined as the 5-day mean divided by the 20-day mean minus one.
Both use the absolute-error loss, at most 1,000 trees with early stopping after 50 rounds, 80\% sampling of rows and inputs, and an L2 penalty of 1.0.
In the screening, we searched 12 settings (number of leaves 31, 63 or 127, learning rate 0.05 or 0.1, and at least 20 or 50 rows per leaf) for each layout and target.
The ensemble is the equal-weight average of LEAR and per-hour LightGBM.

The screening also included three neural networks (LSTM \citep{hochreiter1997long}, a temporal convolutional network \citep{bai2018empirical} and a Transformer \citep{vaswani2017transformer}) for NO1-NO5 and SE1-SE4, each averaged over three seeds.
In the original screening, none was more accurate than per-hour LightGBM on the mean over NO1-NO5, on which the selection was made, or on the mean over SE1-SE4 (Table~S1), so they were not carried forward. In single zones the networks were sometimes more accurate (NO2, NO5, SE3 and SE4).
The screening was run with an earlier carbon series. A rerun with the corrected commodity inputs (Section~\ref{sec:d_data}) left every frozen choice unchanged (Supplementary Section~S2).
The frozen choices were the level-and-scale target for all learned models, per-hour LightGBM with 127 leaves, a learning rate of 0.05 and at least 50 rows per leaf as the main model, and pooled LightGBM with 63 leaves, a learning rate of 0.05 and at least 50 rows per leaf.

\subsection{Accuracy measures and tests}
\label{sec:m_tests}
We measured accuracy by the relative MAE (rMAE), the MAE divided by the MAE of the D-7 naive forecast on the same hours \citep{lago2021}.
The rMAE is comparable across zones with different price levels, and it is not dominated by hours with prices close to zero, which are frequent in the northern zones.
We also report the improvement in MAE over the D-1 naive forecast in per cent, the most intuitive baseline.
The MAE and the root mean squared error (RMSE) are reported in the supplementary material.
All forecasts of a zone were scored on the same hours, and the slot filled on the day of the spring clock change was excluded.

We compared forecasts with the multivariate Diebold--Mariano (DM) test in the form of Lago et al.~\citep{lago2021,diebold1995,ziel2018}.
For two models A and B, the daily loss differential is $\Delta_d = \lVert \varepsilon^{A}_d \rVert_1 - \lVert \varepsilon^{B}_d \rVert_1$, the difference of the sums of absolute errors over the 24 hours of day $d$.
The statistic $\sqrt{N_d}\,\hat\mu/\hat\sigma$ of the mean and standard deviation of $\Delta_d$ is compared with the standard normal distribution in a one-sided test.
Where a conclusion rests on several tests, we adjusted the $p$-values with the Holm procedure \citep{holm1979} and, as a less strict check, reported Benjamini--Hochberg $q$-values \citep{benjamini1995}.

\subsection{Experiments}
\label{sec:m_experiments}
We use out-of-sample predictive content as the criterion. If adding an input improves forecasts in one market type but not in another, the prices of the two carry different information about that input.
The idea follows the predictive notion of Granger \citep{granger1969}, but we do not make causal claims.
The five experiments follow the order of the Results.
All use the monthly rolling scheme of Section~\ref{sec:m_design} and the frozen models unless stated otherwise. The transfer matrix is the one exception to the rolling scheme.

\subsubsection{Price history}
\label{sec:m_history}
We examined the break in the price history in two ways.
The transfer matrix trains pooled LightGBM on a single year $i$ and tests it on a single year $j$, for every pair of years from 2018 to 2025.
It uses fixed settings (63 leaves, a learning rate of 0.05 and at least 20 rows per leaf), since the matrix is a diagnostic rather than a forecasting set-up.
The number of trees for year $i$ is set by early stopping on its last 56 days.
The diagonal cells are estimated by cross-validation within the year, where each quarter is predicted by a model fitted on the other quarters with seven days left out on either side.
We repeated the matrix with the raw, level-only and level-and-scale targets and measured each cell by the rMAE relative to the D-7 naive forecast of year $j$.
The matrix shows how well models carry across price regimes.

The training-window experiment compared per-hour LightGBM trained on the last 365, 730 and 1,095 days, on the expanding window, and on the expanding window without delivery days in 2021-2022.
We also averaged the 365-, 730-, 1,095-day and expanding forecasts with equal weights, without new fits.
To separate the roles of normalisation and history length, we added the raw target with the 365-day and expanding windows, using the raw per-hour setting chosen in the screening (31 leaves, a learning rate of 0.05 and at least 20 rows per leaf).
Each version was tested against the expanding window in both directions, with Holm adjustment within each zone.

\subsubsection{Benchmark}
We compared the two naive forecasts, LEAR, pooled and per-hour LightGBM and the ensemble in all ten zones.
All 30 ordered pairs of the six forecasts were tested in each zone, with Holm adjustment within the zone.

\subsubsection{Neighbouring prices}
\label{sec:m_neighbours}
We added the prices of the previous day of DE-LU, DK1 and NL to per-hour LightGBM for the nine Norwegian and Swedish zones.
These prices are known at the forecast time, because they were set by the auction on D-2.
Each neighbour's 24 prices were normalised with that neighbour's own level and spread, and one level-gap input per neighbour, $(L_{\text{neighbour}} - L_{\text{zone}})/S_{\text{zone}}$, was added, giving 75 new inputs.
The settings were not tuned again.
The improvement in a zone is $\delta$ = rMAE without the neighbouring prices minus rMAE with them, so $\delta > 0$ means that they help.

Our three expectations and their tests were specified before any 2025 forecast was made.
We expected $\delta$ to increase with the coupling measure across the nine zones (H1; one-sided Spearman rank correlation, exact permutation $p$-value over all 362,880 orderings) and no detectable improvement in the four isolated zones (H2; one-sided DM tests, Holm adjustment over the four zones).
We also expected the neighbouring prices to reduce the association of daily errors with the TTF gas price in the four coupled zones (H3), tested by double machine learning (Section~\ref{sec:m_errors}) with the difference of daily MAE without and with the neighbouring prices as the outcome and the gas price as the treatment, with cross-fitting over calendar months and without the holiday dummy, as specified in advance (unlike the fold scheme of Section~\ref{sec:m_errors}), and an inverse-variance weighted one-sided $z$-test over the four zones.

\subsubsection{Value of the TSO forecasts}
We refitted per-hour LightGBM without the TSO forecasts in two versions.
The first removes both the load and the wind and solar forecasts, keeping the prices, the weekday, $L_D$ and $S_D$.
The second removes only the wind and solar forecasts and keeps the load forecasts.
The second version isolates the forecasts whose publication time is uncertain (Section~\ref{sec:infoset}).
Each version was compared with the full model in both directions, with Holm adjustment across the ten zones within each version and direction.
In NO5 the wind and solar forecast is zero in every year, so its forecasts without it are identical to those of the full model.
The further analyses of these forecasts in Section~\ref{sec:r_tso} and Supplementary Sections~S9 and S10 were decided after the 2025 results had been seen and are post hoc. They comprise the relation of the loss to the generation mix of the ten zones, the gain in EUR/MWh and the addition of weekly reservoir inputs to per-hour LightGBM. They also comprise the relation of the daily gain in the Norwegian zones to the deviation of reservoir filling from its seasonal norm, whose hypothesis and tests were written down before it was run, and a repetition of the comparison on 2024, the screening year, as a robustness check. The rerun of the screening with the corrected commodity inputs (Section~\ref{sec:m_models}) was also made after the 2025 results.

\subsubsection{Sources of forecast error}
\label{sec:m_errors}
To ask under which market conditions the main model is less accurate, we related its daily errors to nine conditions describing the market on D-1 and D. The wind and solar forecast is subject to the publication caveat of Section~\ref{sec:infoset}.
The design of this analysis was written down before it was run but was not part of the expectations above.
The outcome is the daily MAE of per-hour LightGBM divided by the zone's 2025 MAE of the D-7 naive forecast, which puts all zones on one scale.
The conditions are the closes of D-2 of TTF gas, EUA carbon (front-December future), API2 coal, Brent oil and EUR/NOK, the reservoir anomaly, the daily means of the wind and solar and of the load forecasts for D, and the temperature at 10:00 on D-1.
The reservoir anomaly is the NVE fill share minus its 20-year median for the same week in Norway \citep{nve_magasin} and the ENTSO-E stored energy minus its same-week mean over 2018-2023 in Sweden.
Each condition was standardised within its zone.

We used the partially linear model of double machine learning \citep{robinson1988,chernozhukov2018}.
For zone $z$ and condition $T$, $Y_d = \theta_z T_d + g(X_d) + u_d$ and $T_d = m(X_d) + v_d$, where the background $X_d$ holds the week of the year (sine and cosine), the weekday, a public-holiday dummy, the change of the mean price of D-1 from D-2, and the other eight conditions.
The holiday dummy covers the official public holidays of the zone's country, plus Christmas Eve and New Year's Eve in Norway and Sweden and Midsummer Eve in Sweden.
A direct version also adds $L_D$ and $S_D$.
The functions $g$ and $m$ were estimated by LightGBM with 7 leaves and 200 trees, cross-fitted over five folds assigned by ISO week number, with one day on each side of every test block left out of training.
The coefficient $\theta_z$ is the slope of the outcome residuals on the condition residuals, with Newey--West standard errors with seven lags \citep{newey1987}.
It measures how much the scaled error changes per standard deviation of the condition after adjustment for the background, and it describes an association, not a causal effect.

To ask whether these associations depend on coupling, we stacked the residuals of the nine zones and regressed the outcome residuals on the condition residuals and on their product with the centred coupling measure.
The main term gives the association at average coupling and the interaction coefficient its change per unit of coupling, and both are reported.
DE-LU is reported separately as a reference and is not part of the pooled regression.
The 95\% intervals and $p$-values of the pooled regression come from a bootstrap that resamples whole days with replacement and keeps all zones of a day together, so that days are treated as clusters (2,000 resamples) \citep{efron1979,cameron2008}.
We adjusted the nine interaction $p$-values with the Holm procedure within each version and re-estimated the interaction nine times, each time without one zone, because it rests on only nine zones.
The wind and solar forecast of NO5 is zero in every year and was left out for that zone.

\subsection{Implementation}
\label{sec:m_impl}
All analyses were run in Python 3.13 with LightGBM 4.7, scikit-learn 1.9, statsmodels 0.15, NumPy 2.5, pandas 3.0, SciPy 1.18, PyTorch 2.14 (for the neural networks in the screening) and entsoe-py 0.8.1 for the data download.
Random seeds were fixed (42 for LightGBM and the nuisance models; 42, 7 and 123 for the neural networks; a fixed seed for every bootstrap).
The models were run on CPU nodes of the Saga cluster of Sigma2, with eight cores per job.
The code and the download scripts are publicly available at \url{https://github.com/myptd/nordic-price-divergence} (see Data and code availability).

\section{Results}
\label{sec:results}

\subsection{Two groups of zones remain separated, with NO5 between them}
\label{sec:r_types}

\begin{figure}[!tbp]
\centering
\includegraphics[width=0.6\textwidth]{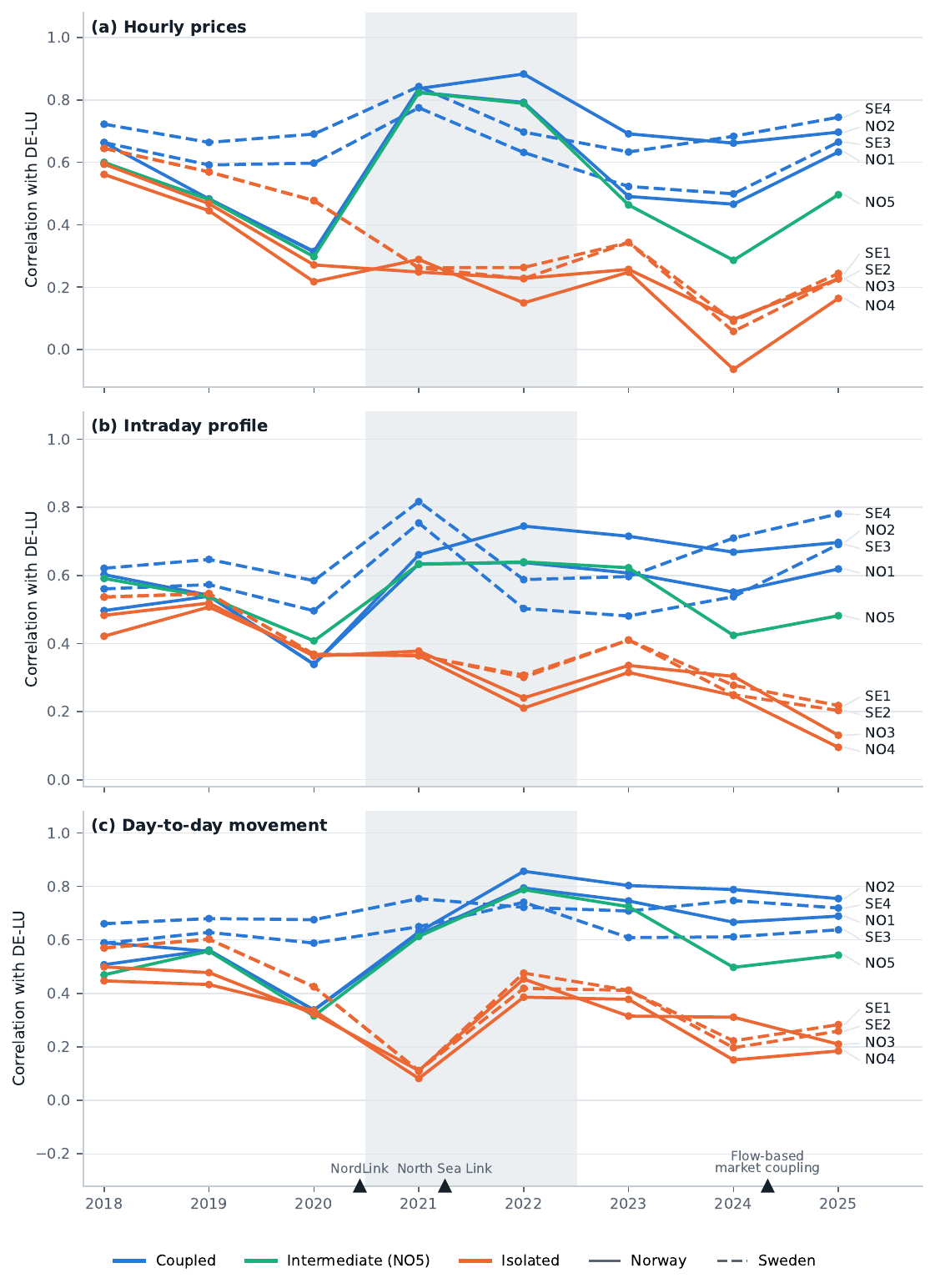}
\caption{Correlation with DE-LU of day-ahead prices by zone and year, 2018-2025.
(a) Hourly prices.
(b) Intraday profile, the hourly price minus the mean price of its day.
(c) Day-to-day movement, the daily mean price minus the mean of its calendar month.
Colours give the market type from the 2024 coupling measure, applied to all years, and line styles give the country.
The shaded years are the crisis years 2021 and 2022.
The triangles in (c) mark the start of trial operation of NordLink (9 December 2020) and North Sea Link (1 October 2021) and the start of Nordic flow-based market coupling (29 October 2024).}
\label{fig:divergence}
\end{figure}

\begin{table}[!tbp]
\centering
\caption{Market types.
Zones are ordered by the 2024 coupling measure within each type.
Generation shares are computed from the ENTSO-E actual generation per production type in 2024 \citep{entsoe_tp}.
Hydro combines run-of-river, reservoir and pumped-storage generation, thermal combines fossil fuels, biomass and waste, and other combines the remaining types. Pumped-storage consumption is excluded.
Swedish thermal generation appears largely under other. We did not verify how Swedish combined heat and power is classified in the ENTSO-E data.
Coupling is the 2024 correlation with DE-LU of the intraday profile (the measure used in all tests) and of the day-to-day movement.
The mean price is the mean of the hourly day-ahead prices in 2025.}
\label{tab:types}
\begin{adjustbox}{max width=\textwidth}
\begin{tabular}{llrrrrrrrrr}
\toprule
 & & \multicolumn{6}{c}{Generation share 2024 (\%)} & \multicolumn{2}{c}{Coupling 2024} & Mean price \\
\cmidrule(lr){3-8} \cmidrule(lr){9-10}
Zone & Country & Hydro & Wind & Solar & Nuclear & Thermal & Other & Intraday & Day-to-day & 2025 (EUR/MWh) \\
\midrule
\multicolumn{11}{l}{\textit{Coupled}} \\
SE4 & Sweden & 16 & 64 & 8 & 0 & 0 & 13 & 0.710 & 0.747 & 60.4 \\
NO2 & Norway & 92 & 8 & 0 & 0 & 0 & 0 & 0.668 & 0.788 & 65.4 \\
NO1 & Norway & 95 & 4 & 0 & 0 & 1 & 0 & 0.551 & 0.666 & 58.4 \\
SE3 & Sweden & 15 & 13 & 1 & 65 & 0 & 5 & 0.538 & 0.611 & 46.2 \\
\midrule
\multicolumn{11}{l}{\textit{Intermediate (NO5)}} \\
NO5 & Norway & 99 & 0 & 0 & 0 & 1 & 0 & 0.424 & 0.497 & 46.9 \\
\midrule
\multicolumn{11}{l}{\textit{Isolated}} \\
NO3 & Norway & 74 & 25 & 0 & 0 & 0 & 1 & 0.304 & 0.311 & 21.1 \\
SE1 & Sweden & 69 & 31 & 0 & 0 & 0 & 1 & 0.277 & 0.223 & 16.7 \\
SE2 & Sweden & 66 & 32 & 0 & 0 & 0 & 2 & 0.249 & 0.197 & 16.5 \\
NO4 & Norway & 79 & 14 & 0 & 0 & 6 & 1 & 0.248 & 0.151 & 8.7 \\
\midrule
\multicolumn{11}{l}{\textit{DE-LU (reference)}} \\
DE-LU & Germany--Luxembourg & 6 & 32 & 14 & 0 & 47 & 1 & -- & -- & 89.3 \\
\bottomrule
\end{tabular}
\end{adjustbox}
\end{table}

Co-movement with DE-LU diverged within both countries, and after the crisis two groups of zones remain clearly separated, with NO5 between them (Fig.~\ref{fig:divergence}).
Until 2019, the nine Nordic zones did not form separate groups, although SE4 always moved most closely with DE-LU.
The year-by-year values are given in Table~S3.

The split opened in Sweden in 2020 and in Norway in 2021.
In 2020, SE3 and SE4 kept their co-movement with DE-LU (0.60 and 0.69), while SE1 and SE2 fell to 0.48.
In 2021, NO1, NO2 and NO5 rose to above 0.8, while NO3 and NO4 stayed below 0.3.
The intraday profile and the day-to-day movement show the same timing.
The Norwegian split coincides with the start of NordLink and North Sea Link and with rising gas prices, whereas the Swedish split came a year earlier, before the gas crisis and without a new cable. This paper does not separate these factors.

After the crisis, the two groups remain clearly separated.
This pattern agrees with Trebbien et al.~\citep{trebbien2024} for the crisis years, and our yearly measures show that it persisted in 2024 and 2025.
In 2024, the intraday coupling measure was 0.538-0.710 in the southern zones SE3, NO1, NO2 and SE4 and 0.248-0.304 in the northern zones NO4, SE2, SE1 and NO3, a gap of 0.234 (Table~\ref{tab:types}).
The gap between the groups is 0.23-0.40 in 2024 and 2025 at all three time scales.
NO5 is the only zone between them. It moved with the southern group in 2021-2023 (intraday 0.62-0.64) and lay between the groups in 2024 (0.424) and 2025 (0.48).
Within the southern group, SE3 and NO1 move less closely with DE-LU than NO2 and SE4.
The correlation of hour-to-hour price changes ranks the zones almost identically (Spearman $\rho = 0.98$), whereas price convergence ranks them differently ($\rho = 0.45$; Table~S4).
Convergence with DE-LU was rare in every zone in 2024 (4-19\% of hours within 1~EUR/MWh), and SE1 and SE2 converged more often than NO1 and NO2, so it measures equal prices rather than shared movement.
In hourly prices, 2024 was the year of the widest spread (NO4 $-0.06$, SE4 0.68).
This drop was only partly a matter of diverging seasonal price levels.
It was smaller in the intraday profile than in hourly prices, but the day-to-day movement of NO5, SE1, SE2 and NO4 also fell by about 0.2 from 2023.
In the Norwegian zones, the drop was concentrated in August-November 2024 (monthly correlations of hourly prices with DE-LU, not shown) and began before the start of flow-based market coupling. A smaller dip occurred in July-October 2023.
In hourly prices, all zones moved closer to DE-LU again in 2025, the first full year of Nordic flow-based market coupling \citep{statnett2024fb}, while in the intraday profile the northern zones moved further away (NO3 from 0.30 to 0.13, NO4 from 0.25 to 0.10).
The yearly clustering of all zones, including DK1, DK2 and NL, is shown in Figs.~S1-S3.

The generation mix forms the second dimension of the market types (Table~\ref{tab:types}).
The Norwegian zones are dominated by hydropower (74-99\% of generation in 2024).
The isolated zones of both countries combine hydropower (66-79\%) with wind power (14-32\%).
The coupled Swedish zones differ from all others. SE3 is dominated by nuclear power (65\%) and SE4 by wind power (64\%).
DE-LU generates mostly from thermal plants (47\%), wind (32\%) and solar (14\%).
Mean prices in 2025 were 46-65~EUR/MWh in the coupled zones and NO5, 9-21~EUR/MWh in the isolated zones and 89~EUR/MWh in DE-LU.

\subsection{One forecasting approach serves all zones}
\label{sec:r_history}
\label{sec:r_benchmark}

\begin{figure}[!tbp]
\centering
\includegraphics[width=\textwidth]{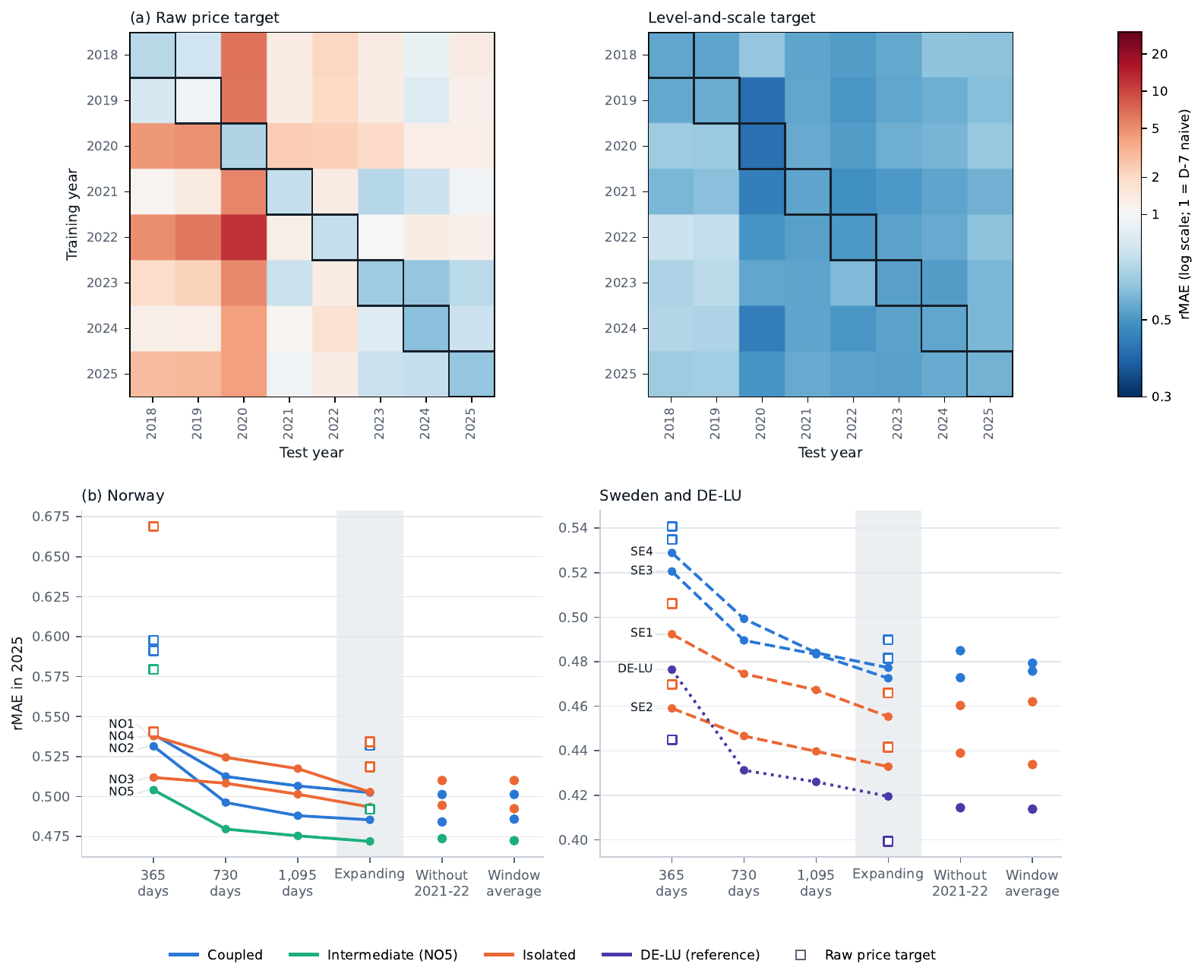}
\caption{Price history and model transfer.
(a) Transfer matrices of pooled LightGBM trained on one year and tested on another, mean rMAE over NO1-NO5, with the raw target (left) and the level-and-scale target (right).
The colour scale is logarithmic and centred at 1, the accuracy of the D-7 naive forecast. Framed cells are the within-year estimates.
The level-only target and the matrices of all zones are shown in Fig.~S4.
(b) rMAE in 2025 of per-hour LightGBM by training window for Norway (left) and for Sweden and DE-LU (right).
Lines connect the windows from 365 days to the expanding window (shaded), the default in all experiments.
The expanding window without 2021-2022 and the equal-weight average of the four windows are shown as separate points.
Open squares show the raw target with the 365-day and the expanding window.}
\label{fig:history}
\end{figure}

Comparing the information in prices across zones requires one forecasting approach that works in all of them, across the break in the price history.
With the raw target, models do not carry over between price regimes, whereas with the level-and-scale target they do (Fig.~\ref{fig:history}a).
With the raw target, 41 of the 56 off-diagonal cells of the Norwegian mean matrix are less accurate than the D-7 naive forecast (mean rMAE 2.17, at worst 11.6 for a model trained on 2022 and tested on 2020), while with the level-and-scale target none is (mean 0.57, at worst 0.77).
The level-only target lies in between (Fig.~S4).

Once prices are normalised, a longer history helps and the crisis years do not harm (Fig.~\ref{fig:history}b).
The mean rMAE over the Norwegian zones falls from 0.525 with 365 days to 0.491 with the expanding window, and the 365-day window is significantly less accurate than the expanding window in 8 of the 10 zones (all but NO3 and NO4).
Leaving out 2021-2022 does not significantly improve accuracy in any zone and is significantly worse in SE4 (Holm $p = 0.030$).
Normalisation helps most when the history is short. It lowers the Norwegian mean rMAE by 0.071 with 365 days and by 0.028 with the expanding window.

\begin{table}[!tbp]
\centering
\caption{Accuracy of the 2025 day-ahead forecasts by market type.
Entries are rMAE, the unitless ratio of the forecast's MAE to the MAE of the D-7 naive forecast on the same hours. For example, 0.761 for naive D-1 in SE4 means that its MAE is 76.1\% of that of naive D-7.
Zones are ordered by the 2024 coupling measure within each type.
Bold marks the forecast with the lowest MAE in each zone.
A dagger marks a forecast that per-hour LightGBM beats significantly (multivariate DM test, Holm adjustment over the 30 pairs within the zone, $p < 0.05$). No forecast beats per-hour LightGBM significantly in any zone.
In SE4 the ensemble and per-hour LightGBM differ only beyond the third decimal.
The bottom rows give the reduction in MAE of per-hour LightGBM relative to the D-1 naive forecast, its MAE in EUR/MWh and the 2025 mean price of the zone.
MAE and RMSE are reported in Table~S2 and all pairwise tests in Fig.~S5.}
\label{tab:benchmark}
\begin{adjustbox}{max width=\textwidth}
\begin{tabular}{lrrrrrrrrrr}
\toprule
 & \multicolumn{4}{c}{Coupled} & \multicolumn{1}{c}{} & \multicolumn{4}{c}{Isolated} & \multicolumn{1}{c}{} \\
\cmidrule(lr){2-5} \cmidrule(lr){7-10}
Forecast & SE4 & NO2 & NO1 & SE3 & NO5 & NO3 & SE1 & SE2 & NO4 & DE-LU \\
\midrule
Naive D-1 & 0.761$^{\dagger}$ & 0.694$^{\dagger}$ & 0.698$^{\dagger}$ & 0.742$^{\dagger}$ & 0.624$^{\dagger}$ & 0.659$^{\dagger}$ & 0.666$^{\dagger}$ & 0.677$^{\dagger}$ & 0.687$^{\dagger}$ & 0.792$^{\dagger}$ \\
Naive D-7 & 1.000$^{\dagger}$ & 1.000$^{\dagger}$ & 1.000$^{\dagger}$ & 1.000$^{\dagger}$ & 1.000$^{\dagger}$ & 1.000$^{\dagger}$ & 1.000$^{\dagger}$ & 1.000$^{\dagger}$ & 1.000$^{\dagger}$ & 1.000$^{\dagger}$ \\
LEAR & 0.520$^{\dagger}$ & 0.525$^{\dagger}$ & 0.536$^{\dagger}$ & 0.508$^{\dagger}$ & 0.497$^{\dagger}$ & 0.527$^{\dagger}$ & 0.484 & 0.483$^{\dagger}$ & 0.528 & 0.508$^{\dagger}$ \\
LightGBM pooled & 0.509$^{\dagger}$ & 0.508$^{\dagger}$ & 0.515 & 0.485 & 0.482 & 0.499 & 0.463 & 0.442 & 0.529 & 0.433 \\
LightGBM per hour & 0.477 & \textbf{0.485} & 0.503 & 0.473 & \textbf{0.472} & \textbf{0.493} & \textbf{0.455} & \textbf{0.433} & \textbf{0.503} & \textbf{0.419} \\
Ensemble & \textbf{0.477} & 0.491 & \textbf{0.502} & \textbf{0.472} & 0.475 & 0.499 & 0.458 & 0.447 & 0.508 & 0.436 \\
\midrule
Improvement over naive D-1 (\%) & 37.3 & 30.0 & 28.0 & 36.3 & 24.4 & 25.1 & 31.7 & 36.1 & 26.9 & 47.0 \\
MAE of LightGBM per hour (EUR/MWh) & 18.2 & 11.8 & 12.5 & 16.1 & 8.5 & 7.9 & 9.6 & 9.5 & 4.3 & 13.8 \\
Mean price 2025 (EUR/MWh) & 60.4 & 65.4 & 58.4 & 46.2 & 46.9 & 21.1 & 16.7 & 16.5 & 8.7 & 89.3 \\
\bottomrule
\end{tabular}
\end{adjustbox}
\end{table}

With this target and the expanding window, per-hour LightGBM is the most accurate forecast, or statistically indistinguishable from it, in every zone (Table~\ref{tab:benchmark}).
No forecast is significantly more accurate than per-hour LightGBM in any zone, and it is significantly more accurate than LEAR in 8 of the 10 zones (all but NO4 and SE1).
It reduces the MAE of the D-1 naive forecast by 24-30\% in Norway, by 32-37\% in Sweden and by 47\% in DE-LU, with an MAE of 4.3-18.2~EUR/MWh in the Nordic zones and 13.8~EUR/MWh in DE-LU.
The smaller gain in Norway reflects a stronger naive benchmark, because the previous day's price is already a better forecast in the Norwegian zones (naive D-1 rMAE 0.62-0.70) than in Sweden (0.67-0.76) and in DE-LU (0.79).
Differences between zones in the following experiments therefore reflect the markets rather than the choice of model.
These MAE values are higher than in our earlier study of the Norwegian zones \citep{phan2026arxiv}, which forecast each price one hour ahead with all prices up to the previous hour, whereas at 11:00 on D-1 the horizon is 13-36 hours and no price after D-1 is known, so the two sets of errors cannot be compared directly.

\subsection{The information in prices does not follow coupling}
\label{sec:r_neighbours}

\begin{figure}[!tbp]
\centering
\includegraphics[width=0.72\textwidth]{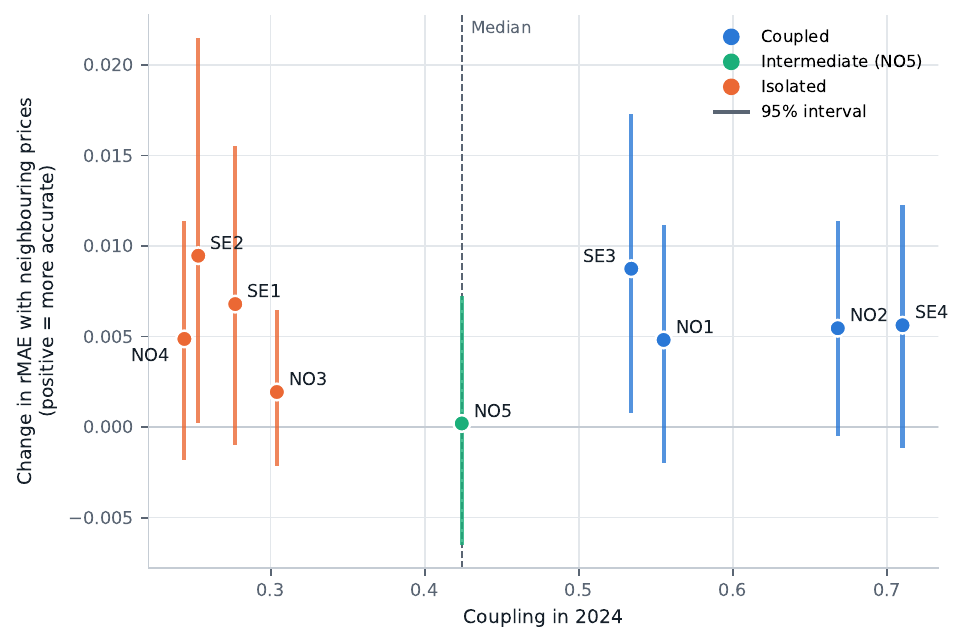}
\caption{Change in 2025 rMAE of per-hour LightGBM when previous-day neighbouring prices (the D-1 prices of DE-LU, DK1 and NL) are added, against the 2024 coupling measure.
Positive values mean that the neighbouring prices improve accuracy.
Vertical lines are 95\% intervals from a bootstrap over days (2,000 resamples).
The dashed line marks the median coupling (NO5), which separates the coupled from the isolated zones.
The largest upper bound is about 0.02 (SE2). The mean gain of the four coupled zones minus that of the four isolated zones is 0.0004 ($-0.0055$ to $0.0061$).
Points of zones with almost equal coupling are shifted slightly apart.}
\label{fig:neighbours}
\end{figure}

Neighbouring prices improve accuracy slightly in every zone except NO5, where the gain is essentially zero, and the gain does not increase with coupling (Fig.~\ref{fig:neighbours}).
The rMAE falls by 0.0002 (NO5) to 0.0095 (SE2), which lowers the MAE by 0.04-2.2\%.
The model without the neighbouring prices is never significantly more accurate.
As a descriptive complement to the tests, the day bootstrap excludes gains larger than about 0.02 in rMAE (under 5\% of MAE) in every zone.

The rank correlation between the gain and coupling is slightly negative ($\rho = -0.10$, $p = 0.61$), and the mean gain in the four coupled zones exceeds that in the four isolated zones by only 0.0004 (95\% day-bootstrap interval $-0.0055$ to $0.0061$).
No low-coupling zone shows a significant gain (Holm $p = 0.140$-$0.179$), but their gains (0.0019-0.0095) are as large as those in the coupled zones (0.0048-0.0088).
Nor do the neighbouring prices reduce the gas-related error in the coupled zones (pooled $\theta = 0.16$, s.e.\ 0.17, $p = 0.170$; NO2 $+0.94$, SE3 $-1.47$). The three expectations specified before the 2025 runs, their tests and outcomes are given in Table~S5.
This pattern is consistent with coupled zones already carrying continental information in their own price history, or with the neighbouring prices carrying mainly region-wide signals (Section~\ref{sec:d_neighbours}).

\subsection{The value of wind and solar forecasts differs between generation systems}
\label{sec:r_tso}

\begin{figure}[!tbp]
\centering
\includegraphics[width=0.85\textwidth]{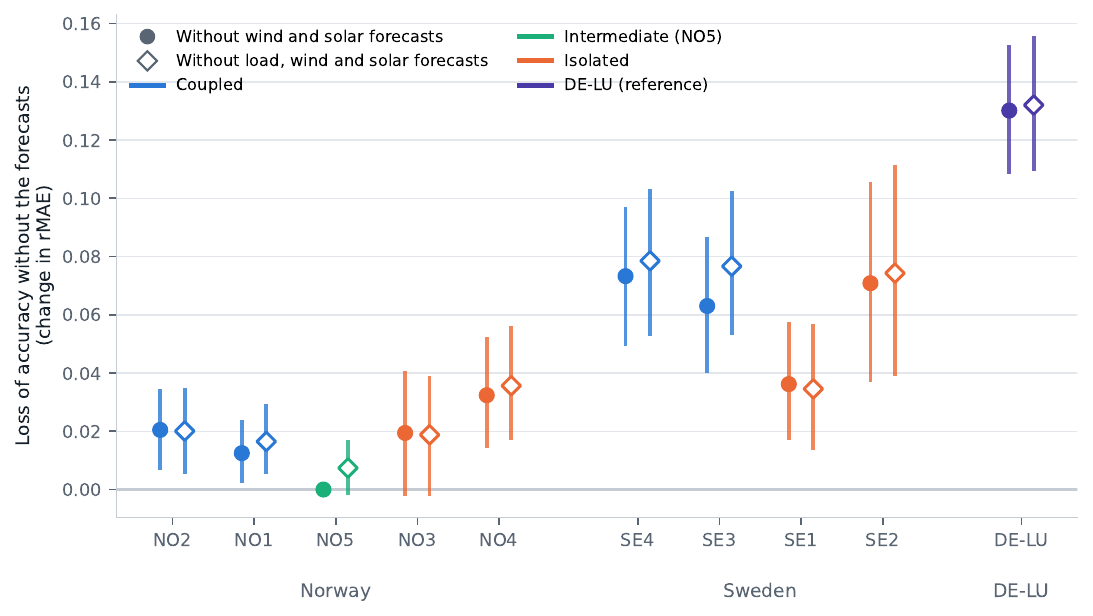}
\caption{Loss of accuracy in 2025 when per-hour LightGBM is fitted without the TSO forecasts, as the change in rMAE against the full model.
Filled circles remove the wind and solar forecasts and keep the load forecasts. Open diamonds remove the load, wind and solar forecasts.
Zones are grouped by country and ordered by the 2024 coupling measure within each country.
Vertical lines are 95\% intervals from a bootstrap over days (2,000 resamples).
NO5 has no wind or solar forecast, so its forecasts without them equal those of the full model.
The losses assume that the forecasts are available at 11:00 on D-1 (Section~\ref{sec:infoset}).}
\label{fig:tso}
\end{figure}

Removing the wind and solar forecasts costs little accuracy in Norway and much more in Sweden and DE-LU (Fig.~\ref{fig:tso}).
The rMAE rises by 0-0.032 in Norway (MAE up to 6.4\% higher), by 0.036-0.073 in Sweden (8.0-16.4\%) and by 0.130 in DE-LU (31\%).
In EUR/MWh, the MAE rises by 0.28-0.50 in NO1-NO4, by 0.77-2.80 in Sweden and by 4.27 in DE-LU (Table~S6).
The loss is significant in all zones except NO3 (Holm $p = 0.062$) and NO5, where the forecasts are identical by construction.
Removing the load forecasts as well adds little (Norwegian mean 0.020 against 0.017 without the wind and solar forecasts only; at most 0.014 more in any zone).

The loss differs between generation systems rather than with coupling.
Within Norway, coupled and isolated zones lose similar amounts (0.012-0.032), and within Sweden the coupled zones SE3 and SE4 and the isolated zone SE2 lose the most (0.063-0.073).
Across the ten zones, the loss increases with the 2024 share of wind and solar generation (Spearman $\rho = 0.89$, exact $p = 0.001$) and decreases with the hydropower share ($\rho = -0.93$, $p < 0.001$; Fig.~S8).
With ten zones, however, the country and the generation mix cannot be separated, because both distinguish Norway from Sweden and DE-LU.
The Norwegian zones also lose less than Swedish zones with similar wind shares (NO3, 25\% wind, 0.019; SE1, 31\% wind, 0.036), although SE1 and SE2 are mainly hydropower as well (66-69\%).
In absolute terms, though, every Swedish zone had a higher peak hourly wind output in 2024 (2.0-5.8~GW) than any Norwegian zone (at most 1.8~GW; Table~S9).
Wind and solar forecasts therefore improve price forecasts little in Norway beyond what the price history contains, and much in Sweden and DE-LU.

The same ordering appears in 2024, the year in which the model settings were chosen (Table~S8).
The ranking of the zones by the gain in EUR/MWh is almost the same in both years (Spearman $\rho = 0.93$), although NO3 gains more in 2024 than in 2025 (0.67 against 0.31~EUR/MWh) and SE3 less (1.10 against 2.14~EUR/MWh).

Reservoir information adds no detectable value at the day-ahead horizon, either as an input or as a condition for the value of wind and solar forecasts.
Weekly ENTSO-E stored energy does not improve per-hour LightGBM in any zone.
We added the reservoir inputs of the pooled model, the latest weekly stored energy complete at 11:00 on D-1 and its change over one and four weeks, without retuning. The NVE fill-share anomaly was left out because its 20-year norm is not known at the forecast time.
The rMAE changes by $-0.003$ to $+0.0003$, with no significant gain or loss in any zone (Holm across the nine zones), and the day bootstrap excludes gains larger than about 0.003 in every zone.
Nor does the daily gain from wind and solar forecasts show a detectable dependence on reservoir filling in the Norwegian zones.
Regressed on the deviation of stored energy from its 2018-2024 median for the same week, with month fixed effects, it shows no negative slope in any zone (Holm $p \geq 0.55$), and the pooled NO1-NO4 slope is $+0.04$~EUR/MWh per standard deviation (s.e.\ 0.23; Table~S7, Fig.~S9).
This does not mean that reservoirs are unimportant for price formation. Their information enters through the price level, which changes slowly.

\subsection{In isolated zones, prices are easier to forecast on days with high forecast wind and solar output}
\label{sec:r_errors}

\begin{figure}[!tbp]
\centering
\includegraphics[width=\textwidth]{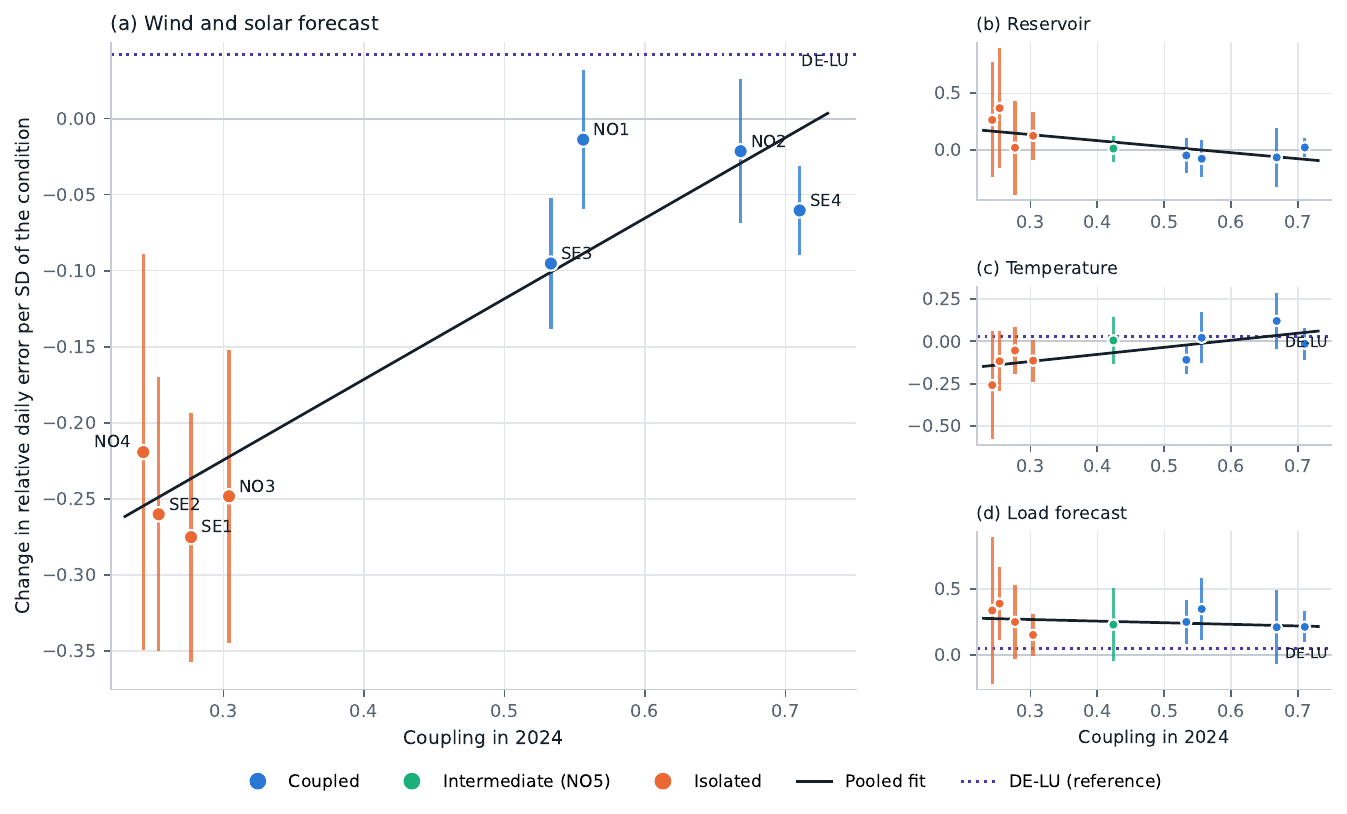}
\caption{Association of the daily errors of per-hour LightGBM in 2025 with market conditions, against the 2024 coupling measure (total version).
(a) Wind and solar forecast.
(b) Reservoir anomaly.
(c) Temperature.
(d) Load forecast.
Points are zone estimates $\theta_z$, the change in the daily MAE relative to the zone's D-7 naive MAE per standard deviation of the condition, with 95\% Newey--West intervals.
The solid line is the pooled fit over the Nordic zones and the dotted line is DE-LU, which is not pooled.
Panels (b)-(d) share the zone positions of panel (a).
NO5 has no wind or solar forecast and DE-LU has no reservoir series.
All nine conditions are shown in Fig.~S6 and the direct version in Fig.~S7.}
\label{fig:errors}
\end{figure}

Prices in the isolated zones are easier to forecast on days with high forecast wind and solar output, while in the coupled zones the day's wind and solar conditions hardly relate to forecast difficulty (Fig.~\ref{fig:errors}).
In the four isolated zones, the relative daily error is lower on days with a high wind and solar forecast ($\theta_z$ from $-0.22$ to $-0.28$ per standard deviation), while in the coupled zones the association is small ($-0.01$ to $-0.10$).
The pooled interaction with coupling is $+0.53$ (95\% interval 0.37-0.72, Holm $p < 0.001$), and it stays between 0.50 and 0.62 when any one zone is left out.
In DE-LU the association is close to zero (0.04).
The isolated zones have substantial local wind power (14-32\% of generation in 2024; Table~\ref{tab:types}).
A possible reading is that on windy days local supply exceeds what the internal bottlenecks let these zones export, so that their prices fall low and flat, and the previous day's prices already anticipate them.
This reading is an interpretation, under which the result concerns local supply and the export limits of the isolated zones rather than continental demand or imports to the continent.

The reservoir and the temperature show the same pattern less clearly.
Errors are larger when reservoirs are fuller than usual in NO3, NO4 and SE2, and the interaction with coupling is $-0.53$ (Holm $p = 0.070$, $q = 0.030$).
For temperature, the interaction is $+0.42$ (Holm $p = 0.032$, $q = 0.018$), with the largest zone estimate in absolute value in NO4 ($-0.26$).
When the level and spread of the previous day's prices are added to the background, the wind and solar interaction remains ($+0.49$, Holm $p < 0.001$), while those of the reservoir and the temperature weaken (Holm $p = 0.55$ and $0.11$).

The load forecast is associated with larger errors in every Nordic zone, independent of coupling (main term 0.25, 95\% interval 0.15-0.36; interaction Holm $p = 1.0$).
No commodity price shows an association that depends on coupling after adjustment.
These estimates describe associations of forecast errors with conditions known on D-1 and D, not causal effects.

\section{Discussion}
\label{sec:discussion}

Since the energy crisis, the Nordic zones have differed in two ways that do not coincide. In how prices move, the coupled south follows the continent and the isolated north does not, and this split has lasted beyond the crisis years. In what prices contain, however, the dividing line runs between countries and generation systems. Neighbouring prices add as little in the south as in the north, whereas wind and solar forecasts add little in Norway and much in Sweden and DE-LU. NO1 and NO2 illustrate the difference, because they move with the continent as closely as SE3, yet their prices gain far less from wind and solar forecasts. We first discuss why the two dimensions can separate, then turn to neighbouring prices, the isolated zones and the price history, and finally to the implications, future work and limitations.

\subsection{Why co-movement and price information separate}

Interconnection and the generation system plausibly act on prices through different channels. An interconnector links a zone's price to the continental price, so that continental shocks such as gas prices and German wind output reach the zone, and co-movement measures this link. The value of a local wind or solar forecast, in contrast, depends on how much local supply varies with the weather and on how much of that variation reaches the price. A zone can therefore be closely coupled to the continent while its own wind and solar output matters little for its price, as in NO1 and NO2, which generate 92-95\% of their electricity from hydropower (Table~\ref{tab:types}).

Norwegian prices are also more persistent than Swedish and continental prices. Repeating the previous day's price is already a stronger forecast in Norway (naive D-1 rMAE 0.62-0.70) than in Sweden (0.67-0.76) or DE-LU (0.79), which leaves less room for any input to improve accuracy (Table~\ref{tab:benchmark}). Reservoir information behaves in the same way. Weekly stored energy does not improve the forecasts, and the association between reservoir conditions and forecast errors weakens when the previous day's price level is added to the error diagnostics (Section~\ref{sec:r_errors}). Reservoir information therefore appears to enter prices through their slowly changing level rather than to be unimportant for price formation.

One possible explanation for the small value of wind and solar forecasts in Norway is the flexibility of reservoir hydropower. Producers with large reservoirs can shift generation over time and thereby absorb local fluctuations in wind and solar output before they reach the price. Two results limit this explanation. First, within the Norwegian zones the value of the forecasts showed no detectable dependence on reservoir filling in 2025, so if flexibility matters, it acts as a structural property of the system rather than through the current filling level. Second, SE1 and SE2 also generate two thirds of their electricity from hydropower, yet their prices gain more from wind and solar forecasts than Norwegian prices do. Their wind output is also larger in absolute terms, with peak hourly wind generation of 2.8 and 5.8~GW in 2024 against at most 1.8~GW in any Norwegian zone (Table~S9). The hydropower share alone therefore cannot explain the differences, and the amount of wind and solar power in the local supply appears to matter as well. With ten zones, however, the country and the generation mix cannot be separated, so these readings remain interpretations rather than tested mechanisms.

\subsection{Neighbouring prices}
\label{sec:d_neighbours}
Previous-day neighbouring prices add little beyond each zone's own price history, and they add about as much in isolated as in coupled zones (Section~\ref{sec:r_neighbours}). Our result complements Mascarenhas et al.~\citep{mascarenhas2026}, who improve SE3 forecasts by 9\% with prices that earlier-closing markets have already published for the delivery day. Read together, the two results suggest that cross-border information is valuable when it describes the delivery day itself and largely redundant when it describes the previous day, which the zone's own price history already reflects. This is our interpretation of the two studies, not a result we tested. Our result also contrasts with the gains from market integration in Lago et al.~\citep{lago2018integration}, who studied other markets and periods. Chomon and Ziel~\citep{chomon2026} cover the same Nordic and continental zones and a test period that overlaps ours, and they find that networked models outperform single-zone models. Their network effects, however, come from the neighbours' residual load forecasts for the delivery day rather than from neighbouring prices. This agrees with our reading that cross-border information helps when it describes the delivery day.

Two readings fit our result. The continental information may already be contained in the coupled zones' own prices, so that the neighbouring prices repeat it. Alternatively, the prices of D-1 may carry mainly region-wide signals, such as a general change of level, that help all zones alike. The experiment cannot separate the two readings, but it does bound the size of the effect. Gains larger than about 0.02 in rMAE are excluded in every zone, and the gain in coupled zones is at most about 0.006 larger than in isolated zones (Section~\ref{sec:r_neighbours}).

Our small, uniform gain agrees with Bendiksen et al.~\citep{bendiksen2025}, whose in-sample association between cross-border flows and the tails of the NO2 price distribution does not translate into improved out-of-sample forecasts of that distribution. Although we use prices and point forecasts and they use flows and quantile forecasts, the two studies point the same way. Their tail result suggests that continental information may matter mainly in extreme hours, which point forecasts scored by the MAE do not reward. Bj{\o}rndal et al.~\citep{bjorndal2026} show that NordLink increased NO2 price volatility and more than doubled the sensitivity of NO2 prices to European gas prices. In our experiments, the gas price is not associated with the scaled forecast error, neither at average coupling nor as a function of coupling (Fig.~S6). This is consistent with the level-and-scale normalisation (Eq.~\ref{eq:norm}) absorbing much of that exposure, although we did not test it.

\subsection{Isolated zones and local supply}
In the error diagnostics, prices in the isolated zones are easier to forecast on days with high forecast wind and solar output, while in the coupled zones the day's wind and solar conditions hardly relate to forecast difficulty (Section~\ref{sec:r_errors}). The isolated zones have substantial local wind power, 14-32\% of their generation in 2024 (Table~\ref{tab:types}). One possible reading is that on windy days local supply exceeds what the internal bottlenecks let these zones export, so that their prices fall low and flat and are easier to forecast. This result does not contradict the small value of wind and solar forecasts as inputs in Norway, because the two results measure different things. An input can add little to average accuracy while the condition it describes still marks easy and hard days, for example when windy days bring low, flat prices that the previous day's prices already anticipate. If wind capacity in the northern zones continues to grow, local wind conditions could matter more for their prices, which is an implication to be tested rather than a finding.

\subsection{The price history}
The break in the price history can be bridged in two ways that act as partial substitutes (Section~\ref{sec:r_history}). With the level-and-scale target, a longer history still helps and the crisis years do not harm accuracy, whereas with the raw target a longer history helps more but the model remains significantly less accurate in NO1-NO4. This agrees with the finding that the best calibration window varies across markets and periods \citep{marcjasz2018calibration} and with the value of averaging across windows \citep{hubicka2019}. Unlike the good results of short windows for boosting models in volatile markets \citep{michalakopoulos2025}, however, the 365-day window was significantly less accurate than the expanding window in 8 of the 10 zones once prices were normalised. The normalisation plays the role of the variance-stabilising transformations used for linear models \citep{uniejewski2018,uniejewski2026}, and it also lets tree-based models, which cannot predict outside the range of their training targets, learn from years with very different price levels.

\subsection{Practical implications}

Our results have three implications for system operators, market participants and policymakers. First, the value of wind and solar forecasts for day-ahead prices differs strongly between generation systems. If they were available before gate closure, they would lower the MAE of the price forecast by 0.28-0.50~EUR/MWh in NO1-NO4, by 0.77-2.80~EUR/MWh in Sweden and by 4.27~EUR/MWh in DE-LU. Since European rules require TSOs to publish these forecasts only by 18:00 on D-1, after the auction has closed \citep{eu543}, earlier publication would matter most in zones with much wind and solar power and little in hydropower-dominated Norway.

Second, the choice of forecasting inputs should follow the generation system rather than the degree of coupling. Previous-day neighbouring prices add little even in the zones most closely coupled to the continent, whereas local wind and solar forecasts are worth acquiring in Sweden and DE-LU. For Norwegian hydropower producers, a forecast based mainly on the zone's own price history and load forecasts captures most of the information available from our inputs.

Third, the co-movement of each zone with continental prices is a measurable property of the market that has persisted beyond the crisis years rather than a passing effect of the crisis. Because it indicates how closely a zone's prices follow continental prices, it is relevant to hedging and to the debate on new interconnectors, whereas it says little about which information is useful for forecasting. In addition, one modelling approach, per-hour LightGBM with a normalised target and an expanding window, served all zone types, so that crisis-period data can stay in the training set and the same method can be used across markets.

\subsection{Future work}
\label{sec:d_future}
Five directions follow from our results and limitations.
\begin{itemize}
\item Separating the causes of the split, namely the new cables, the crisis and hydrology, for example with an event study around the start of NordLink and North Sea Link and with Sweden as a contrast.
\item Separating the country from the generation mix, for example with more zones, installed-capacity data or the growth of wind power in the northern zones over time.
\item Testing forward-looking neighbour information, such as prices already published by earlier-closing markets for the delivery day \citep{mascarenhas2026} and neighbouring wind, solar and load forecasts for the delivery day, together with networked models \citep{chomon2026}.
\item Probabilistic forecasts, to test whether continental information matters mainly in extreme hours, as the tail results of Bendiksen et al.~\citep{bendiksen2025} suggest.
\item TSO forecasts with recorded publication times and the effect of Nordic flow-based market coupling, once several years of post-2024 data are available.
\end{itemize}

\subsection{Limitations}
\label{sec:d_limits}

Several limitations qualify our findings. First, the evaluation covers one test year, 2025, with 2024 as a robustness year, and relatively few zones within each market type. The results may therefore depend on the market conditions of these years and may not generalise to other periods or zones. Moreover, our coupling measure captures price co-movement rather than its causes, so we cannot attribute the divergence to interconnectors, transmission bottlenecks or the energy crisis.

Second, the relations of forecast value and forecast errors to the generation mix and to market conditions are associations, not causal effects. With ten zones, the country and the generation mix cannot be separated, and the reservoir test covers a single year in which each Norwegian zone saw a different range of reservoir filling. In addition, we assume that wind and solar forecasts are available at 11:00 on D-1, although they may be published as late as 18:00, and that each weekly reservoir value is available from the end of its week, although it is published some days later. The estimated gains therefore measure the value the forecasts would have if published before gate closure, not their value under current publication rules.

Third, the evaluation considers only point forecasts and backward-looking information from neighbouring markets (Section~\ref{sec:d_neighbours}). It therefore does not establish whether neighbouring prices improve probabilistic forecasts, the prediction of extreme price events or forecasts that use information available closer to delivery. Our neighbouring-market inputs also exclude Great Britain, whose day-ahead prices are not available on the ENTSO-E Transparency Platform after Brexit. This omission may matter, because Bendiksen et al.~\citep{bendiksen2025} find that flows with Great Britain and Germany-Luxembourg, unlike those with the Netherlands and Denmark, have the potential to increase price volatility in Norway.

\section{Conclusion}
\label{sec:conclusion}

We asked whether the divergence of Nordic electricity prices after the energy crisis also divides the information that these prices carry, and the answer is no. The split in co-movement is real and lasting, yet coupled and isolated zones hardly differ in how much the previous day's neighbouring prices tell about tomorrow's price. Conversely, zones that move alike can differ several-fold in what local wind and solar forecasts tell about it. NO2 and SE4 follow continental prices about equally closely and had similar mean prices in 2025, but the forecasts improved the price forecast by 0.50~EUR/MWh in NO2 and by 2.80~EUR/MWh in SE4. A zone's coupling to the continent is therefore a poor guide to its information needs, and its generation system is a better one.

This distinction is likely to grow in importance, because interconnection and wind and solar power are both expanding in the Nordic region and act on prices through different channels. Whether it also holds for extreme hours, for information that becomes available closer to delivery and for power systems outside the Nordic region remains to be tested.

\section*{CRediT authorship contribution statement}
\textbf{My Thi Diem Phan:} Conceptualization, Data curation, Methodology, Software, Formal analysis, Visualization, Writing -- original draft.
\textbf{Tuyen Trung Truong:} Data curation, Validation, Writing -- review \& editing.
\textbf{Hoai Phuong Ha:} Data curation, Validation, Writing -- review \& editing.
\textbf{Dat Thanh Nguyen:} Supervision, Methodology, Software, Validation, Writing -- review \& editing.

\section*{Data and code availability}
The code and the scripts that download every input series and produce every table and figure are publicly available at \url{https://github.com/myptd/nordic-price-divergence} under the MIT licence. The repository also contains the weather data, the model forecasts without the observed prices and all result tables and figures. The price, load, generation and reservoir data from ENTSO-E and NVE and the commodity prices from Yahoo Finance and Ember are publicly accessible but are not redistributed because of their licence terms. They can be downloaded with the scripts provided.

\section*{Funding}
This research did not receive any specific grant from funding agencies in the public, commercial, or not-for-profit sectors.

\section*{Declaration of competing interest}
The authors declare no competing interests.

\section*{Declaration of generative AI and AI-assisted technologies in the manuscript preparation process}
During the preparation of this work the authors used AI-assisted language tools (Claude and ChatGPT) to assist with language refinement. After using these tools, the authors reviewed and edited the content as needed and take full responsibility for the content of the manuscript.

\section*{Acknowledgements}
The computations were performed using resources provided by Sigma2, the National Infrastructure for High-Performance Computing and Data Storage in Norway, under projects NN9780K, NS9342K, and NN9342K.

\bibliographystyle{elsarticle-num}
\bibliography{references}

\end{document}